\documentclass[12pt,a4paper]{article}
\makeatletter
\def\eqnarray{%
\stepcounter{equation}%
\let\@currentlabel=\theequation
\global\@eqnswtrue
\global\@eqcnt\z@
\tabskip\@centering
\let\\=\@eqncr
$$\halign to \displaywidth\bgroup\@eqnsel\hskip\@centering
$\displaystyle\tabskip\z@{##}$&\global\@eqcnt\@ne
\hfil$\displaystyle{{}##{}}$\hfil
&\global\@eqcnt\tw@$\displaystyle\tabskip\z@{##}$\hfil
\tabskip\@centering&\llap{##}\tabskip\z@\cr}
\makeatother
\newcommand{\ket}[1]{{\vert{#1}\rangle}}
\newcommand{\bra}[1]{{\langle{#1}\vert}}

\begin{document}

\title{\sl Two--Level System and Some Approximate Solutions 
in the Strong Coupling Regime}
\author{
  Kazuyuki FUJII
  \thanks{E-mail address : fujii@yokohama-cu.ac.jp }\\
  Department of Mathematical Sciences\\
  Yokohama City University\\
  Yokohama, 236-0027\\
  Japan
  }
\date{}
\maketitle
%\thispagestyle{empty}
%
%
%  gaiyou
%
%
\begin{abstract}
  In this paper we treat the 2--level system interacting with external 
  fields without the rotating wave approximation and construct some 
  approximate solutions in the strong coupling regime. 
\end{abstract}

%\newpage

%
%
%     Honbun
%
%
In this paper we consider the $2$--level system interacting with external 
fields (a radiation field for example ) and treat this system without 
the rotating wave approximation (RWA). 
Usually we assume the RWA and solve the equation. This approximation is 
appropriate in the weak coupling regime, however in the strong coupling one 
it is not applicable. See for example \cite{AE} and \cite{AC}. 
How can we treat the system in the strong coupling regime ? 
Shahriar et al treated this problem in \cite{SPM}, \cite{MsS} and \cite{SP} 
and constructed some approximate solutions which contained what they called 
the Bloch--Siegert oscillation. See also \cite{CEC}, \cite{BaWr}, \cite{SGD}, 
\cite{MFr1} as another approachs. 

By the way we have treated some important models coming from quantum optics 
in the strong coupling regime, see \cite{MFr2}, \cite{KF1} and \cite{KF2} 
(also \cite{NPT} for an experiment in the strong coupling regime). 
See \cite{MFr3} as a general introduction in this field. 
This method is also applicable in the case, so we construct some approximate 
solutions which are completely different from those of Shahriar et al. 
We believe that our solutions are more general. 

Let $\{\sigma_{1}, \sigma_{2}, \sigma_{3}\}$ be Pauli matrices : 
\begin{equation}
\sigma_{1} = 
\left(
  \begin{array}{cc}
    0& 1 \\
    1& 0
  \end{array}
\right), \quad 
\sigma_{2} = 
\left(
  \begin{array}{cc}
    0& -i \\
    i& 0
  \end{array}
\right), \quad 
\sigma_{3} = 
\left(
  \begin{array}{cc}
    1& 0 \\
    0& -1
  \end{array}
\right), 
\end{equation}
and $\sigma_{+} = (1/2)(\sigma_{1}+i\sigma_{1})$, 
$\sigma_{-} = (1/2)(\sigma_{1}-i\sigma_{1})$.

We consider an atom with two energy levels which interacts with external 
field with $g\mbox{cos}(\omega t)$. 
The Hamiltonian in the dipole approximation is given by 
\begin{equation}
\label{eq:hamiltonian}
H=-\frac{\Delta}{2}\sigma_{3}+g\ \mbox{cos}(\omega t)\sigma_{1}, 
\quad \mbox{cos}(\omega t)=(1/2)(\mbox{e}^{i\omega t}+\mbox{e}^{-i\omega t}), 
\end{equation}
where $\omega$ is the frequency of the external field, $\Delta$ 
the energy difference between two levels of the atom, $g$ the coupling 
constant between the external field and the atom. 
In the following we cannot assume the rotating wave approximation 
(which neglects the fast oscillating terms), namely the Hamiltonian given by 
\begin{equation}
\label{eq:subhamiltonian}
H=-\frac{\Delta}{2}\sigma_{3}+\frac{g}{2}\left(\mbox{e}^{i\omega t}\sigma_{+}
+\mbox{e}^{-i\omega t}\sigma_{-}\right). 
\end{equation}

\par \noindent 
Mysteriously enough we cannot solve this simple model completely (maybe 
non--integrable), nevertheless we have found this model has a very rich 
structure. 

We would like to solve the Schr{\" o}dinger equation with the Hamiltonian 
(\ref{eq:hamiltonian}) in {\bf the strong coupling regime} $g \gg \Delta$. 
Therefore we change in (\ref{eq:hamiltonian}) a role of 
the kinetic term and the interaction term like 
\[
\label{eq:duality}
H = H_{0}-\frac{\Delta}{2}\sigma_{3} 
  \equiv g\ \mbox{cos}(\omega t)\sigma_{1}-\frac{\Delta}{2}\sigma_{3}. 
\]
First we solve the interaction term which is an easy task. Before that 
let us make some preliminaries. 

Let $W$ be a Walsh--Hadamard matrix 
\begin{equation}
W=\frac{1}{\sqrt{2}}
\left(
  \begin{array}{cc}
    1& 1 \\
    1& -1
  \end{array}
\right)
=W^{-1}
\end{equation}
then we can diagonalize $\sigma_{1}$ by using this $W$ as 
$
\sigma_{1}=W\sigma_{3}W^{-1}
$.
The eigenvalues of $\sigma_{1}$ is $\{1,-1\}$ with eigenvectors 
\[
\ket{1}=\frac{1}{\sqrt{2}}
\left(
  \begin{array}{c}
    1 \\
    1
  \end{array}
\right), \quad 
\ket{-1}=\frac{1}{\sqrt{2}}
\left(
  \begin{array}{c}
    1 \\
    -1
  \end{array}
\right). 
\]
We note that $\sigma_{3}\ket{1}=\ket{-1},\ \sigma_{3}\ket{-1}=\ket{1}$ and 
\begin{eqnarray}
\ket{1}\bra{1}&=&\frac{1}{2}
\left(
  \begin{array}{cc}
    1& 1 \\
    1& 1
  \end{array}
\right)=
W
\left(
  \begin{array}{cc}
    1& \\
     & 0
  \end{array}
\right)
W
=
W(|0)(0|)W, \nonumber \\
\ket{-1}\bra{-1}&=&\frac{1}{2}
\left(
  \begin{array}{cc}
    1& -1 \\
    -1& 1
  \end{array}
\right)=
W
\left(
  \begin{array}{cc}
    0& \\
     & 1
  \end{array}
\right)
W
=
W(|1)(1|)W, \nonumber 
\end{eqnarray}
where $|0)=(1, 0)^{\mbox{t}}$ and $|1)=(0, 1)^{\mbox{t}}$. 

\par \noindent 
Then it is easy to see 
\begin{equation}
H_{0}
=g\mbox{cos}(\omega t)\ket{1}\bra{1}-g\mbox{cos}(\omega t)\ket{-1}\bra{-1}, 
\end{equation}
so the Schr{\" o}dinger equation 
$
i\frac{d}{dt}\psi=H_{0}\psi
$
can be solved to be  
\begin{equation}
\label{eq:partialsolution}
\psi(t)
 =\left\{
 \mbox{e}^{-i \frac{g}{\omega} sin(\omega t)} \ket{1}\bra{1} +
 \mbox{e}^{i \frac{g}{\omega} sin(\omega t)} \ket{-1}\bra{-1} 
 \right\}\psi_{0}, 
\end{equation}
where $\psi_{0}$ is a constant vector. 
Since we want to solve the full Schr{\" o}dinger equation 
\begin{equation}
\label{eq:fullequation}
i\frac{d}{dt}\psi=\left(H_{0}-\frac{\Delta}{2}\sigma_{3}\right)\psi, 
\end{equation}
we appeal to the {\bf method of constant variation}. Namely under 
$
\psi_{0} \longrightarrow \psi_{0}(t)
$ 
in (\ref{eq:partialsolution}) we substitute (\ref{eq:partialsolution}) into 
(\ref{eq:fullequation}), then we obtain 
\begin{eqnarray}
\label{eq:}
&&i\frac{d}{dt}\psi_{0}=-\frac{\Delta}{2}\times  \nonumber \\
&&\quad 
\left\{
 \mbox{e}^{i \frac{g}{\omega} sin(\omega t)} \ket{1}\bra{1} +
 \mbox{e}^{-i \frac{g}{\omega} sin(\omega t)} \ket{-1}\bra{-1} 
\right\}\sigma_{3}
\left\{
 \mbox{e}^{-i \frac{g}{\omega} sin(\omega t)} \ket{1}\bra{1} +
 \mbox{e}^{i \frac{g}{\omega} sin(\omega t)} \ket{-1}\bra{-1} 
\right\}
\psi_{0} \nonumber \\
&&\qquad \quad =-\frac{\Delta}{2}
\left\{
 \mbox{e}^{2i \frac{g}{\omega} sin(\omega t)} \ket{1}\bra{-1} +
 \mbox{e}^{-2i \frac{g}{\omega} sin(\omega t)} \ket{-1}\bra{1} 
\right\}\psi_{0} \nonumber \\
&&\qquad \quad =-\frac{\Delta}{2}
W
\left\{
 \mbox{e}^{2i \frac{g}{\omega} sin(\omega t)}|0)(1| +
 \mbox{e}^{-2i \frac{g}{\omega} sin(\omega t)}|1)(0|  
\right\}W\psi_{0}, \nonumber 
\end{eqnarray}
so we have only to solve the equation 
\begin{equation}
\label{eq:subequation}
i\frac{d}{dt}\phi=-\frac{\Delta}{2}
\left\{
 \mbox{e}^{2i \frac{g}{\omega} sin(\omega t)}\sigma_{+} +
 \mbox{e}^{-2i \frac{g}{\omega} sin(\omega t)}\sigma_{-}
\right\}\phi, \quad \phi\equiv W\psi_{0}
\end{equation}
where $|0)(1|=\sigma_{+}$ and $|1)(0|=\sigma_{-}$. 

Here we note that the solution $\psi$ in (\ref{eq:fullequation}) is given 
as follows 
\begin{equation}
\label{eq:full-solution}
\psi(t)
=W\left\{
 \mbox{e}^{-i \frac{g}{\omega} sin(\omega t)}|0)(0| +
 \mbox{e}^{i \frac{g}{\omega} sin(\omega t)}|1)(1|
 \right\}\phi(t)
\end{equation}
with $\phi$ in (\ref{eq:subequation}). 

To solve the equation (\ref{eq:subequation}) we set : 
\begin{equation}
\label{eq:ansatz}
\phi=
\left(
  \begin{array}{c}
    a \\
    b
  \end{array}
\right).
\end{equation}

\par \noindent 
By substituting (\ref{eq:ansatz}) into (\ref{eq:subequation}) we obtain 
the integral equations 
\begin{eqnarray}
a(t)&=&
\alpha+i\frac{\Delta}{2}
\int_{0}^{t}\mbox{e}^{2i \frac{g}{\omega} sin(\omega s)}b(s)ds
, \\
b(t)&=&
\beta+i\frac{\Delta}{2}
\int_{0}^{t}\mbox{e}^{-2i \frac{g}{\omega} sin(\omega s)}a(s)ds
,
\end{eqnarray}
or 
\begin{eqnarray}
a(t)&=&
\alpha+i\frac{\Delta}{2}\beta
\int_{0}^{t}\mbox{e}^{2i \frac{g}{\omega} sin(\omega s)}ds-
\frac{\Delta^{2}}{4}
\int_{0}^{t}\mbox{e}^{2i \frac{g}{\omega} sin(\omega s)}
\left\{
\int_{0}^{s}\mbox{e}^{-2i \frac{g}{\omega} sin(\omega x)}a(x)dx
\right\}
ds
, \\
b(t)&=&
\beta+i\frac{\Delta}{2}\alpha
\int_{0}^{t}\mbox{e}^{-2i \frac{g}{\omega} sin(\omega s)}ds-
\frac{\Delta^{2}}{4}
\int_{0}^{t}\mbox{e}^{-2i \frac{g}{\omega} sin(\omega s)}
\left\{
\int_{0}^{s}\mbox{e}^{2i \frac{g}{\omega} sin(\omega x)}b(x)dx
\right\}
ds
,
\end{eqnarray}
where $\alpha,\ \beta$ are integral constants. 
Therefore in the lowest order with respect to $\Delta$,\ 
$\phi$ in (\ref{eq:ansatz}) is given by 
\[
\phi=
\left(
  \begin{array}{c}
    \alpha+i\frac{\Delta}{2}\beta
    \int_{0}^{t}\mbox{e}^{2i \frac{g}{\omega} sin(\omega s)}ds \\
    \beta+i\frac{\Delta}{2}\alpha
    \int_{0}^{t}\mbox{e}^{-2i \frac{g}{\omega} sin(\omega s)}ds
  \end{array}
\right).
\]

From here we consider the physical condition. $\psi$ in 
(\ref{eq:full-solution}), therefore $\phi$ must be normalized (
$\phi^{\dagger}\phi=1$) in the lowest order, so we can set 
$|\alpha|^{2}+|\beta|^{2}=1$. 
From (\ref{eq:full-solution}) the approximate solutions 
that we are looking for are 
\begin{equation}
\label{eq:approximate-solution}
\psi(t)=\frac{1}{\sqrt{2}}
\left(
  \begin{array}{cc}
    1& 1 \\
    1& -1
  \end{array}
\right)
\left(
  \begin{array}{c}
  \mbox{e}^{-i \frac{g}{\omega} sin(\omega t)}
  \left\{
    \alpha+i\frac{\Delta}{2}\beta
    \int_{0}^{t}\mbox{e}^{2i \frac{g}{\omega} sin(\omega s)}ds 
  \right\}\\
  \mbox{e}^{i \frac{g}{\omega} sin(\omega t)}
  \left\{
    \beta+i\frac{\Delta}{2}\alpha
    \int_{0}^{t}\mbox{e}^{-2i \frac{g}{\omega} sin(\omega s)}ds
  \right\}
  \end{array}
\right), 
\end{equation}
where $|\alpha|^{2}+|\beta|^{2}=1$.

Finally we make a comment : If we write $\psi$ above as 
\[
\psi=
\left(
  \begin{array}{c}
    \psi_{0}\\
    \psi_{1}
  \end{array}
\right),
\]
then we have 
\begin{equation}
|\psi_{1}|^{2}=\frac{1}{2}
\left(
1-\mbox{cos}\left(\frac{2g}{\omega}\mbox{sin}(\omega t)\right)
\right) 
\end{equation}
under (B) if $\alpha=\beta=\mbox{e}^{i\theta}/\sqrt{2}$. 

Our solutions (\ref{eq:approximate-solution}) are more or less well--known. 
However our (simple) method will become a starting point to find more 
complicated approximate solutions or to generalize the model from 2--levels 
to n--levels, \cite{KF3}. 

\par \vspace{5mm} 
The author would like to Marco Frasca for several useful comments.

%%%%%%%%%%%%%
%References%
%%%%%%%%%%%%%


\begin{thebibliography}{99}
%
\bibitem{AE}L. Allen and J. H. Eberly : 
\newblock Optical Resonance and Two--Level Atoms, 
\newblock Wiley, New York, 1975. 
%
\bibitem{AC}A. Corney : 
\newblock Atomic and Laser Spectroscopy, 
\newblock Oxford University Press, 1977. 
%
\bibitem{SPM}M. S. Shahriar, P. Pradhan and J. Morzinski : 
\newblock Determination of the phase of an electromagnetic field 
via incoherent detection of fluorescence, 
\newblock quant-ph/0205120.
%
\bibitem{MsS}M. S. Shahriar : 
\newblock Frequency Locking Via Phase Mapping of Remote Clocks Using 
Quantum Entanglement, 
\newblock quant-ph/0209064. 
%
\bibitem{SP}M. S. Shahriar and P. Pradhan : 
\newblock Fundamental limitation on qubit operations due to 
the Bloch--Siegert Oscillation, 
\newblock in the Proceedings of the Quantum Communication, Measurement 
and Computing (QCMC'02), 
\newblock quant-ph/0212121.
%
\bibitem{CEC}C. E. Creffield : 
\newblock Location of crossing in the spectrum of a driven two--level 
system, 
\newblock cond-mat/0301168. 
%
\bibitem{BaWr}J. C. A. Barata and W. F. Wreszinski : 
\newblock Strong Coupling Theory of Two Level Atoms in Periodic Fields, 
\newblock Phys. Rev. Lett. 84(2000), 2112, 
\newblock physics/9906029. 
%
\bibitem{SGD}A. Santana, J. M. Gomez Llorente and V. Delgado : 
\newblock Semiclassical dressed states of two-level quantum systems driven 
by nonresonant and/or strong laser fields, 
\newblock J. Phys. B 34(2001), 2371, 
\newblock quant-ph/0011015.
%
\bibitem{MFr1}M. Frasca : Theory of dressed states in quantum optics, 
\newblock Phys. Rev. A 60(1999), 573, 
\newblock quant-ph/9811037. 
%
\bibitem{MFr2}M. Frasca : 
\newblock Rabi oscillations and macroscopic quantum superposition states, 
\newblock Phys. Rev. A 66(2002), 023810,  
\newblock quant-ph/0111134. 
%
\bibitem{MFr3}M. Frasca : A modern review of the two--level approximation, 
\newblock quant-ph/0209056. 
%
\bibitem{KF1} K. Fujii : 
\newblock Mathematical Structure of Rabi Oscillations in the Strong 
Coupling Regime, 
\newblock to appear in J. Phys. A, 
\newblock quant-ph/0203135. 
%
\bibitem{KF2} K. Fujii : 
\newblock N--Level System Interacting with Single Radiation Mode and 
Multi Cat States of Schr{\" o}dinger in the Strong Coupling Regime, 
\newblock quant-ph/0210166. 
%
\bibitem{KF3} K. Fujii et al : 
\newblock N--Level System and Some Approximate Solutions 
in the Strong Coupling Regime (tentative), 
\newblock in progress. 
%
\bibitem{NPT} Y. Nakamura, Yu. A. Pashkin and J. S. Tsai : 
\newblock Rabi Oscillations in a Josephson--Junction Charge Two--Level 
System, 
\newblock Phys. Rev. Lett., 87(2001), 246601.
%
\end{thebibliography}
\end{document}